\documentstyle[epsfig,aps,floats,preprint]{revtex}

\newcommand{\beq}{\begin{equation}}
\newcommand{\eeq}{\end{equation}}
\newcommand{\bea}{\begin{eqnarray}}
\newcommand{\eea}{\end{eqnarray}}

\def\laq{\raise 0.4ex\hbox{$<$}\kern -0.8em\lower 0.62
ex\hbox{$\sim$}}
\def\gaq{\raise 0.4ex\hbox{$>$}\kern -0.7em\lower 0.62
ex\hbox{$\sim$}}

\def \pa {\partial}

\def \b {\beta}
\def \a {\alpha}

\def \ga {\gamma}

\def \r {\rho}

\begin{document}
\par
\begingroup

\begin{flushright}
BA-TH/99-331\\
March 1999\\
gr-qc/9905062\\
\end{flushright}

\vspace{12mm}
{\large\bf\centering\ignorespaces
Looking back in time beyond the big bang
\vskip2.5pt}

\bigskip
{\dimen0=-\prevdepth \advance\dimen0 by23pt
\nointerlineskip \rm\centering
\vrule height\dimen0 width0pt\relax\ignorespaces
M. Gasperini
\par}

{\small\it\centering\ignorespaces
Dipartimento di Fisica, Universit\`a di Bari, \\
Via Amendola 173, 70126 Bari, Italy \\
and \\Istituto Nazionale di Fisica Nucleare, Sezione di Bari,
Bari, Italy \\
\par}

\par
\bgroup
\leftskip=0.10753\textwidth \rightskip\leftskip
\dimen0=-\prevdepth \advance\dimen0 by17.5pt \nointerlineskip
\small\vrule width 0pt height\dimen0 \relax

\vskip -1 cm
\begin{abstract}
String theory can (in principle) describe gravity at all curvature
scales, and can be applied to cosmology to look back in time beyond
the Planck epoch. The duality symmetries of string theory suggest a
cosmological picture in which the imprint of a primordial, pre-big bang
phase could still be accessible to present observations. The predictive
power of such a scenario relies, however, on our ability to connect in a
smooth way the pre-big bang to the present cosmological regime.
Classical radiation back reaction seems to play a key role to this
purpose, by isotropizing and turning into a final expansion any state of
anisotropic contraction possibly emerging from the pre-big bang at
the string scale. 
\end{abstract}

\begin{center}
---------------------------------------------\\
{\sl Essay written for the 1999 
Awards of the Gravity Research Foundation,}\\
{\sl and selected for Honorable Mention.}\\
To appear in {\bf Mod. Phys. Lett. A}
\end{center}

\par\egroup
\thispagestyle{plain}
\endgroup

\pacs{}


The standard cosmological model is, rightfully, one of the most 
celebrated scientific conquests of the present century. Such a  model,
however, cannot be extrapolated back in time beyond an initial regime
approaching a state  of infinite density and curvature -- the so-called
``big bang".  The history of the Universe from the big bang down to the
present time  is more or less well known, and its various aspects are
under active  study since more than forty years \cite{1}. But  what
happened before  the big bang? 

This question has not been raised  until very recently, mainly because
of the  lack of a systematic application to cosmology of the powerful 
instruments of modern theoretical physics, able (in principle) to look 
back in time beyond the Planck scale. As a  consequence, the big bang
has  represented so far a sort of ``Hercule's Pillars " of cosmology. In
the ancient  times, when nobody knew the world beyond the Straits of
Gibraltar,  because nobody sailed the sea beyond that point, it was
common opinion that Gibraltar would represent the end of the world
itself.  In the same way, today, the  big bang is often popularly referred
as the beginning of the Universe,  the beginning of spacetime, the
beginning of ``Everything", just in view  of the lack of  information 
about earlier time scales. There are  also respectable scientific attempt,
in a quantum cosmology context, to explain the  origin of the Universe
and of the spacetime itself as a process of  tunnelling ``from nothing",
i.e. from some unspecified vacuum \cite{2}.  They are affected,
however, by problems of boundary conditions \cite{3},  arising just
because of the ignorance, intrinsic to standard cosmology,  about the
state of the Universe before it emerged at the Planck scale. 

The standard cosmological scenario has been complemented and 
improved,  in many aspects, by the inflationary scenario \cite{4}.
Concerning however the very beginning of the Universe, i.e. the  state
and the evolution of the Universe before the Planck epoch, the 
situation in conventional inflation is not so much different
from that of  the standard model, because a phase of conventional
inflationary  expansion, at constant curvature, cannot be extended back
in time for  ever \cite{5}.  Quoting Alan Guth's recent survey of
inflationary cosmology  \cite{6}:


{\sl ``... Nevertheless, since inflation appears to be eternal 
only into the future, 
but not to the past, an important question remains open. How did 
all start? Although eternal inflation pushes this question far into the 
past, and well beyond the range of observational tests, the question 
does not disappear."}


String theory seems to suggest an answer to this question 
and, most  important, seems to suggest that the beginning of the
Universe, namely its  evolution at times earlier than Planckian, might be
not completely  beyond the range of present observational tests, in
contrast to the  sentence quoted above. 
 The technical instrument used by string theory to look back in time, 
beyond the Planck scale and the big bang singularity, is (a general 
version of) the duality symmetry which, together with 
supersymmetry, is  probably one of the most powerful and important
tools of modern  theoretical physics (at least, because they are both
at the grounds of  superstring theory \cite{7}, which is at present one
of the best candidate for a  Theory of Everything). 

Just like supersymmetry associates to any bosonic state a fermionic
partner, and vice-versa, duality associates to any cosmological
configuration with decreasing curvature a geometric partner with
growing curvature, and vice-versa. Just like supersymmetry
cancellations can eliminate the field theory divergences, duality
symmetries are expected to regularize the spacetime and curvature
singularities. The assumption of (at least approximate)
``self-duality" symmetry, which combines duality and time
reversal, suggests in particular a complete model of
cosmological evolution, defined in cosmic time from minus to plus
infinity, in which the Universe expands around a fixed point of maximal
(finite) curvature, controlled by the fundamental length scale $L_s$ of
string theory \cite{8}.  

The big bang singularity is replaced in this context by a phase of high
(nearly Planckian) curvature, which marks the transition from an initial
accelerated growth of the curvature $H$ and of the string coupling
$g_s$ (parametrized by the dilaton $\phi$ as $g_s=e^{\phi/2}$), to a
final state of radiation-dominated, decelerated expansion at
constant dilaton. It comes natural, in this context, to call ``pre-big
bang" the initial phase of growing curvature,  in contrast to the
subsequent, standard ``post-big bang" evolution, with decreasing
curvature. 

The most revolutionary aspect of this scenario is probably the fact
that the high-curvature, Planckian regime is reached {\em at the end},
and not {\em at the beginning} of inflation. Thus, the state of the
Universe at the Planck scale does not represent an initial condition,
but is rather the result of a long and classical pre-big bang (i.e.
pre-Planckian) evolution, which starts from a state of very low
curvature and small coupling ($H \ll L_s^{-1}$, $ g_s \ll 1$), and is
well controlled by the low-energy string effective action.  In other
words, the Universe is far from being a ``new-born baby" at the time of 
the big bang transition, being instead almost in the middle of a very 
long, possibly infinite, life. 

From a phenomenological point of view, the  important aspect of
this scenario is the fact that the cosmological evolution preceding the
Planck epoch may become accessible to present (direct or indirect)
observations. I would like to recall, in particular, three possible
effects,  referring to observations to be performed {\sl i)}
in a not so  far future, {\sl ii)} in a near future, and {\sl iii)} to
observations already (in  part) performed.  They are, respectively: the
presence of a graviton background much stronger than expected  in
standard  inflation \cite{9}, the contribution of massless (or massive)
axion  fluctuations to the CMB anisotropy spectrum \cite{10}, and the
production of primordial ``seeds" for the cosmic magnetic fields
\cite{11}. 

The predictive power of this scenario relies however on the
construction of non-singular models, describing a smooth transition
from the pre- to the post-big bang regime. Implementing such a
transition is in general problematic in the context of the tree-level,
gravi-dilaton string effective action; there are ``no-go theorems"
\cite{12} excluding a regular transition also in the presence of perfect
fluid and axionic Kalb-Ramond sources, and suggesting the need for
higher order (quantum loops \cite{13} and higher curvature \cite{14})
corrections. Examples of a complete transition through the strong
coupling regime have been implemented, up to date, but only with the
help of ``ad hoc" corrections: a non-local two-loop \cite{15} or
four-loop \cite{16} dilaton potential, a higher derivative dilaton kinetic
term \cite{17}. 

The above non-go theorems are all formulated in the context of
homogeneous and isotropic backgrounds. It is known, on the other
hand, that the singularity can be ``boosted away" already at the
tree-level \cite{18} (through an appropriate transformation of
the global, pseudo-orthogonal duality symmetry group), provided the
metric is allowed to be anisotropic. There are examples \cite{18}, dating
back to the early studies of the pre-big bang scenario, of anisotropic
solutions with a non-trivial axion background which satisfy all the
conditions \cite{17} necessary for a ``graceful exit" from the pre-big
bang phase, and which describe indeed a perfectly smooth transition
from an initial growing curvature and dilaton phase, to a final
decreasing curvature and dilaton phase.

Such examples are usually regarded as unrealistic, mainly because in
the final post-big bang regime the metric background may be
contracting;  if expanding, it is nevertheless highly anisotropic,
with only two dynamical (spatial) dimensions (the background is frozen
in all the other space directions). 

It should be taken into account, however, that the background
transition described by the above solutions generates a large amount
of radiation: the quantum fluctuations of the initial pre-big bang state
are amplified by the accelerated evolution of the background, and
re-enter the horizon in the subsequent decelerated phase,
contributing eventually to the post-big bang sources as a gas of
relativistic particles. In the post-big bang phase, this 
radiation tends to become dominant with respect to the axion sources
\cite{19}, and it is well known that the radiation can isotropize an
initially anisotropic metric \cite{20}.  In a contracting background, we
may expect that the radiation energy density become dominant even
faster, and may even turn the initial  contraction into a final expansion,
as suggested by the general radiation-dominated solution of the
gravi-dilaton cosmological equations \cite{21}.  

To confirm this expectation, I will now present the results of a
numerical computation which shows the changes induced by the
radiation back reaction in the final state of the regular solutions
\cite{18}. The aim is twofold: {\sl 1)} to stress the possibility of a
smooth connection between the pre-big bang Universe and the present
isotropic, expanding Universe also to lowest order in the string
effective action, without any ``ad hoc" higher order correction (the
importance of the low energy string effective action has been recently
stressed also in the context of M-theory \cite{22}); {\sl 2)} to point out
the possible relevance of contracting backgrounds for the solution of
the graceful exit problem of string cosmology. 

I will concentrate on the particular class of regular backgrounds
obtained by boosting the dual of the two-dimensional vacuum Milne
solution. The dilaton,  the non-vanishing components of the metric
and of the antisymmetric axion field, $B_{\mu\nu}=-B_{\nu\mu}$, can
be written in the synchronous gauge as follows \cite{18}:
\bea
&&
\phi= \phi_0 -\ln \left(\b+\a b^2t^2\right), ~~~
g_{11}= -{\a +\b b^2t^2\over \b+\a b^2t^2}, ~~~
g_{12}= -{\sqrt{\a \b }\left(1+b^2t^2\right)\over \b+\a b^2t^2},
 \nonumber\\
&&
g_{00}=-g_{22}=-g_{33}= 1, ~~~  B_{12}=g_{12},  ~~~
\a=\cosh \ga +1, ~~~ \b=\cosh \ga -1,
\label{1}
\eea
where $\phi_0$, $b$ and $\ga$ are real arbitrary parameters. This
background represents an exact anisotropic solution of the tree-level
string cosmology equations: 
\bea
&&
R_\mu\,^\nu +\nabla_\mu\nabla^\nu \phi- 
{1\over 4} H_{\mu\a
\b}H^{\nu\a\b} = 0, ~~~~~
R-(\nabla_\mu\phi)^2+2 \nabla_\mu
\nabla^\mu \phi 
-{1\over
12}H_{\mu\nu\a}H^{\mu\nu\a}=0, \nonumber\\
&&
\pa_\nu\left(\sqrt{|g|}e^{-\phi}H^{\nu\a\b}\right)=0, 
~~~~~~~~~~~~~~~~~~~~H_{\nu\a\b}= 3! \pa_{[\nu}B_{\a\b]}. 
\label{2}
\eea
The time evolution of the dilaton and of $H_1$, which represents the
rate-of-change of the distance along the $x_1$ direction between two
comoving geodesics, is illustrated in Fig. 1. The parameter  $H_1$, which
is the analog of the Hubble parameter for anisotropic, off-diagonal
metrics,  is defined by $H_1=
\theta_{\mu\nu}n^\mu n^\nu$, where $n^\mu$ is a unit space-like
vector along $x_1$, and, in the synchronous frame, 
$\theta_{\mu\nu}=\nabla_{(\mu}u_{\nu)}$ is the so-called expansion
tensor  for a congruence of co-moving geodesics $u^\mu$ \cite{23}. As
clearly shown in Fig. 1, the initial accelerated expansion ($H_1>0, \dot
H_1 >0$), evolves smoothly into a final decelerated contraction ($H_1<0,
\dot H_1 >0$). The evolution is non-trivial only in the $\{x_1,x_2\}$
plane, as $H_2=H_3=0$. Notice that, 
with an appropriate choice of the parameters, it is always possible to
bound the peak values  to be smaller than one in string units, 
consistently with the  low-energy effective action.

\begin{figure}[t]
\begin{center}
\mbox{\epsfig{file=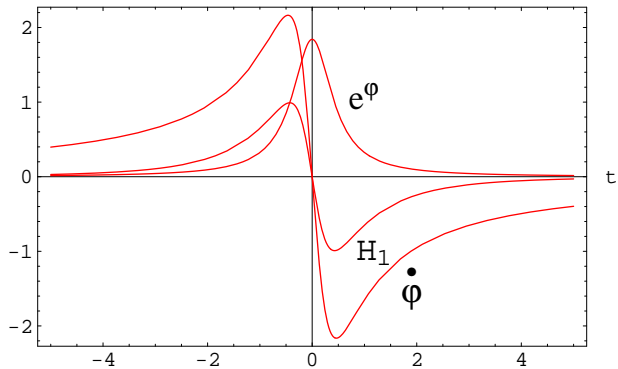,width=82mm}}
\vskip 5mm
\caption{\sl Smooth evolution from an expanding pre-big bang
configuration to a contracting post-big bang configuration, according
to the solution (\ref{1}), with $\phi_0=0$, and $b=\ga=1$.} 
\end{center}
\end{figure}

In the above background, the transition to the post-big bang regime
amplifies the quantum fluctuations of the initial pre-big bang state. In
other words, the final post-big bang state is characterized by a large
number of massless particles (gravitons, dilatons, photons...), produced
in pairs from the vacuum \cite{19}: their total energy density $\r$ is
bounded by the maximal curvature scale of the background, which, in
its turn, is controlled by the string length scale $L_s$. Their effective
averaged stress tensor is traceless \cite{24}, and we can thus
represent their contribution to the post-big bang background like that
of an effective radiation fluid, with $\langle \r\rangle=\langle
3p\rangle$ and $\langle \r\rangle~ \laq ~ L_s^{-4}$. This contribution is
weighed by the dilaton, in the string frame \cite{8}, as  $\langle
\r\rangle e^\phi$, and it is initially subdominant  at the beginning of
the post-big bang phase, but tends to grow in time with respect to the
axion. 

To take into account this back reaction, I have added to the right hand
side of the first of equations (\ref{2}) the contribution of the effective
radiation stress tensor, $e^\phi \langle T_\mu~^\nu\rangle $ (in units
$8 \pi G =1$), and I have numerically integrated the system of
equations (\ref{2}) plus the conservation equation  $\nabla_\nu
\langle T_\mu~^\nu\rangle=0$ (which is still valid in the usual form, in
spite of the dilaton \cite{8}). I have imposed the boundary conditions
that the background starts initially (at large and negative times) in the
configuration described by the solution (\ref{1}), and that the radiation
keeps negligible until the background is well inside the post-big bang
regime. The evolution is thus unchanged in the pre-big bang phase, but
the final stage of the post-big bang evolution is qualitatively affected
by the radiation back reaction, as illustrated in the three following
figures where the results of the numerical integration (plotted as
solid curves, with time measured in units of $b^{-1}$) are compared
to the unperturbed solution. 

\begin{figure}[h]
\begin{center}
\mbox{\epsfig{file=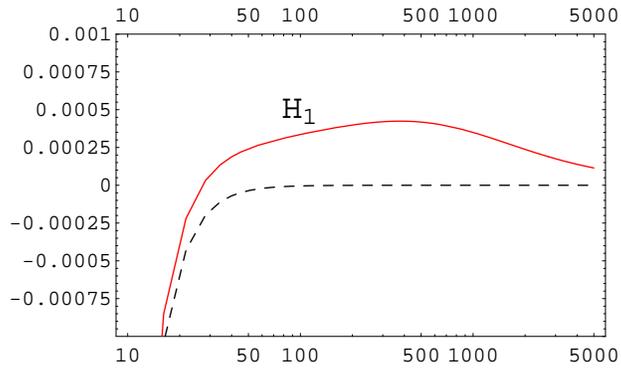,width=82mm}}
\vskip 5mm
\caption{\sl Time evolution of $H_1$. With the inclusion of the
radiation back reaction (solid curve) the decelerated contraction of Fig.
1 (dashed curve) becomes decelerated expansion, with $H_1$ positive
and asymptotically decreasing.}  
\end{center}
\end{figure}

\begin{figure}[h]
\begin{center}
\mbox{\epsfig{file=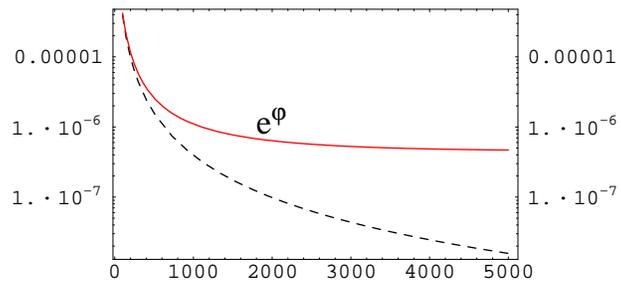,width=82mm}}
\vskip 5mm
\caption{\sl Time evolution of the dilaton. The friction of the radiation
back reaction tends to stop the dilaton (solid curve), with respect to the
axion-dominated solution of Fig. 1 (dashed curve).} 
\end{center}
\end{figure}

\begin{figure}[h]
\begin{center}
\mbox{\epsfig{file=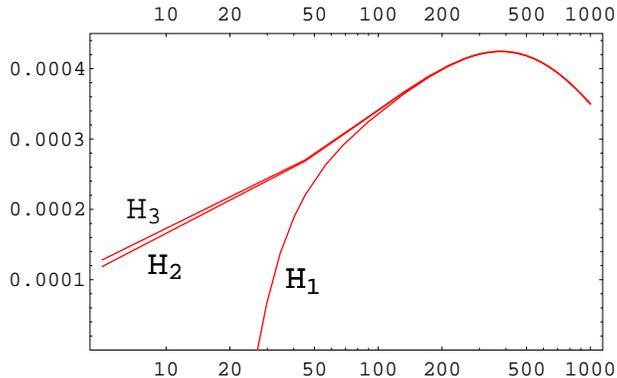,width=82mm}}
\vskip 5mm
\caption{\sl Time evolution of $H_1, H_2, H_3$, with the inclusion of the
radiation back reaction. All the spatial dimensions become dynamical,
and the background converges to a state with the same rate of
decelerated expansion along any direction.} 
\end{center}
\end{figure}

There are three main effects: the contraction turns eventually into a
standard decelerated expansion, with $H_1>0$ and $\dot H_1<0$ (Fig. 2);
the dilatons tends to stop (Fig. 3), as the background converges towards the
radiation-dominated, frozen-dilaton asymptotic solution; 
the frozen spatial dimensions start to expand ($H_1,H_2  \not=0$), and
the expansion tends to become isotropic (Fig. 4), asymptotically
approaching a state in which $H_1=H_2=H_3$, and in which the
expansion rates along the three spatial directions are all positive and
decreasing.

This example is not completely realistic, for various reasons (for
instance, an appropriate non-perturbative dilaton potential is expected
to be included, in the post-big bang phase, to give a mass to the dilaton,
and to fix the final string coupling to a realistic value \cite{25} $\langle
g_s^2\rangle=\langle e^\phi\rangle \sim 10^{-2} - 10^{-4}$). 
Already from this simple example we can  learn, however, that the
back reaction of the produced radiation is possibly a key missing
ingredient in previous studies of the graceful exit problem. It is a
physical effect, not a term added ``ad hoc" to the action, which could 
represent the last step of a a complete transition from the string
perturbative vacuum to the present cosmological state. When such  a
back reaction is included, in particular, it seems possible to have a look
at the pre-big bang Universe even following the geodesics of the
low-energy string effective action.

\acknowledgments
I wish to thank Gabriele Veneziano for many useful discussions. Special 
thanks are also due to Egidio Scrimieri for his precious advice in
improving the plots of the numerical solutions presented in this paper.


\begin{references}
\newcommand{\bb}{\bibitem}

\bb{1} S. Weinberg, {\sl Gravitation and cosmology} (Wiley, New York,
1972);  E. W. Kolb and M. S. Turner, {\sl The Early Universe}, (Addison 
Wesley, Redwood City, Ca, 1990). 

\bb{2}A. Vilenkin, Phys. Rev. D {\bf 30}, 509 (1984); 
A. D. Linde, Sov. Phys. JEPT {\bf 60}, 211 (1984); 
Y. Zel'dovich and A. A. Starobinski, Sov. 
Astron. Lett. {\bf 10}, 135 (1984); V. A. Rubakov, Phys. Lett. B {\bf 
148}, 280 (1984).

\bb{3}J. B. Hartle and S. W. Hawking, Phys. Rev. D {\bf 28}, 2960 
(1983); S. W. Hawking, Nucl. Phys. B {\bf 239}, 257 (1984); S. W. 
Hawking and D. N. Page, Nucl. Phys. B {\bf 264}, 185 (1986).

\bb{4}A. Guth, Phys. Rev. D {\bf 23}, 347 (1981).

\bb{5}A. Vilenkin, Phys. Rev. D {\bf 46}, 2355 (1992); A. Borde, A. 
Vilenkin, Phys. Rev. Lett. {\bf 72}, 3305 (1994).

\bb{6}A. Guth, {\sl The inflationary Universe}, (Vintage, London, 1998). 

\bb{7}M. B. Green, J. Schwartz and E. Witten, {\sl Superstring theory}, 
(Cambridge U. Press, Cambridge, Ma, 1987). 

\bb{8}G. Veneziano, Phys. Lett. B {\bf 265}, 287 (1991); M. Gasperini and
G. Veneziano, Astropart. Phys. {\bf 1}, 317 (1993); Mod. Phys. Lett. A {\bf
8}, 3701 (1993); Phys. Rev. D  {\bf 50}, 2519 (1994). An updated
collections of papers on the pre-big bang  scenario is available at {\tt
http://www.to.infn.it/\~{}gasperin}.

\bb{9}M. Gasperini and M. Giovannini, Phys. Lett. B {\bf 282}, 36 (1992);
Phys. Rev.  D {\bf 47}, 1519 (1993);  R. Brustein, M. Gasperini, M.
Giovannini and G. Veneziano, Phys.  Lett. B {\bf 361}, 45 (1995).

\bb{10}E. J. Copeland, R. Easther and D. Wands,  Phys. Rev. D {\bf 56},
874 (1997); R. Durrer, \\M . Gasperini,
M. Sakellariadou and G. Veneziano, Phys. Lett. B {\bf 436}, 66 (1998); 
Phys. Rev.  D {\bf 59}, 043511 (1999);  M. Gasperini and G. Veneziano, 
Phys. Rev. D {\bf 59}, 043503 (1999).

\bb{11}M. Gasperini, M. Giovannini and G. Veneziano, Phys. Rev. Lett. 
{\bf 75}, 3796 (1995). 

\bb{12}R. Brustein and G. Veneziano, Phys. Lett. B {\bf 329},  429 (1994); 
N. Kaloper, R. Madden and K. A. Olive, Nucl. Phys. B {\bf 452},  677 (1995);
R. Easther, K. Maeda and D. Wands, Phys. Rev. D {\bf 53}, 4247 (1996). 

\bb{13}I. Antoniadis, J. Rizos and K. Tamvakis, Nucl. Phys. B {\bf 415}, 
497 (1994); S. J. Rey, Phys. Rev. Lett. {\bf 77}, 1929 (1996);  M. Gasperini
and G. Veneziano, Phys. Lett. B {\bf 387},  715 (1996); S. Foffa,  M.
Maggiore and R. Sturani,  {\sl Loop corrections and graceful exit in
string cosmology},  IFUP-TH 8/99 (February 1999).  

\bb{14}M. Gasperini, M. Maggiore and G. Veneziano, Nucl. Phys. B {\bf
494},  315 (1997); M. Maggiore, Nucl. Phys. B {\bf 525}, 413 (1998); 
R. Brandenberger, R. Easther and J. Maia, JHEP
{\bf 9808}, 007 (1998). 

\bb{15} Second paper in Ref. [8]. 

\bb{16}M. Gasperini, J. Maharana and G. Veneziano, Nucl. Phys. B {\bf
472},  349 (1996).

\bb{17}R. Brustein and R. Madden, Phys. Lett. B. {\bf 410}, 110
(1997); Phys. Rev. D {\bf 57}, 712 (1998).

\bb{18}M. Gasperini, J. Maharana and G. Veneziano, Phys. Lett. B {\bf
272},  277 (1991). 

\bb{19}M. Gasperini, in {\sl String gravity and
physics at the Planck energy scale}, edited by N. S\'anchez and A.
Zichichi (Kluwer A. P. , Dordrecht, 1996), p. 305.

\bb{20}J. B. Zeldovich and I. D. Novikov, {\sl Relativistic Astrophysics}, 
(Chicago University Press, Chicago, 1983). 

\bb{21}Third paper in Ref. [8]. 

\bb{22}T. Banks, W. Fischler and L. Motl, {\sl Dualities versus
singularities}, hep-th/9811194. 

\bb{23}R. M. Wald, {\sl General relativity},  (Chicago University Press,
Chicago, 1984). 

\bb{24}R. L. Abramo, R. H. Brandenberger and V. F. Mukhanov,   Phys.
Rev. D {\bf 56}, 3248 (1997).

\bb{25}V. Kaplunovsky, Phys. Rev. Lett. {\bf 55}, 1036 (1985). 

\end{references}
\end{document}